\documentclass[twocolumn,aps,pra,amsmath,amssymb]{revtex4-1}
\usepackage[latin9]{inputenc}
\usepackage{amstext}
\usepackage{amssymb}
\usepackage{graphicx}
\usepackage{esint}

\makeatletter
\usepackage{epsfig}
\usepackage{bm}
\usepackage{dcolumn}
\usepackage{color}

\makeatother

\begin{document}

\title{Quantum fluctuations in a strongly interacting Bardeen-Cooper-Schrieffer
polariton condensate at thermal equilibrium}

\author{Hui Hu and Xia-Ji Liu}

\affiliation{Centre for Quantum and Optical Science, Swinburne University of Technology,
Melbourne, Victoria 3122, Australia}

\date{\today}
\begin{abstract}
Microcavity electron-hole-photon systems in two-dimensions are long
anticipated to exhibit a crossover from Bose-Einstein condensate (BEC)
to Bardeen-Cooper-Schrieffer (BCS) superfluid, when carrier density
is tuned to reach the Mott transition density. Yet, theoretical understanding
of such a BEC-BCS crossover largely relies on the mean-field framework
and the nature of the carriers at the crossover remains unclear to
some extent. Here, motivated by the recent demonstration of a BCS
polariton laser {[}Hu \textit{et al.}, arXiv:1902.00142{]} and based
on a simplified short-range description of the electron-hole attraction,
we examine the role of quantum fluctuations in an exciton-polariton
condensate at thermal equilibrium and determine the number of different
type carriers at the crossover beyond mean-field. Near Mott density
and with ultra-strong light-matter coupling, we find an unexpectedly
large phase window for a strongly correlated BCS polariton condensate,
where both fermionic Bogoliubov quasi-particles and bosonic excitons
are significantly populated and strongly couple to photons. We predict
its photoluminescence spectra and show that the upper polariton energy
gets notably renormalized, giving rise to a high-energy side-peak
at large carrier density, as observed in recent experiments.
\end{abstract}

\pacs{71.36.+c, 03.75.Hh, 67.10.-j, 74.78.Na}

\maketitle
Over the past two decades, the realization of quantum fluids of electron-hole-photon
condensates in semiconductor microcavities \cite{Deng2010,Carusotto2013,Byrnes2014,Schneider2017}
has opened a new era for better photonic technologies \cite{Fraser2016,Sanvitto2016}.
In most situations with low carrier density, tightly bound electron-hole
($e$-$h$) pairs can be well approximated as structureless point-like
bosons of small Bohr radius $a_{B}$. Coupled with photons, they turn
into quasi-particles (i.e., exciton-polaritons) and undergo Bose-Einstein
condensation (BEC) at sufficiently low temperatures \cite{Deng2010,Carusotto2013,Byrnes2014,Schneider2017}.
When carrier density is high, comparable to a characteristic Mott
transition density $n_{\textrm{mott}}\sim a_{B}^{-2}$, the composite
nature of $e$-$h$ pairs becomes important and the emerging fermionic
degree of freedoms may eventually lead to a Bardeen-Cooper-Schrieffer
(BCS) polariton condensate \cite{Keeling2007}, where the loosely-bound
fermionic $e$-$h$ pairs are formed by photon-mediated attractions
\cite{Ohashi2002,Citrin2003,Liu2015}, rather than Coulomb interactions
that turn out to be screened. This high-density regime was recently
investigated in several experiments \cite{Horikiri2016,Horikiri2017}.
While the strong coupling between $e$-$h$ pairs and photons was
confirmed via the observation of a negative excitation branch \cite{Horikiri2017}
and a puzzling high-energy side-peak \cite{Horikiri2016} in photoluminescence,
the existence of a BCS polariton condensate remains elusive. In the
highly non-equilibrium lasing regime, the signature of a BCS polariton
laser was mostly recently demonstrated \cite{Hu2019arXiv}.

\begin{figure}
\centering{}\includegraphics[width=0.5\textwidth]{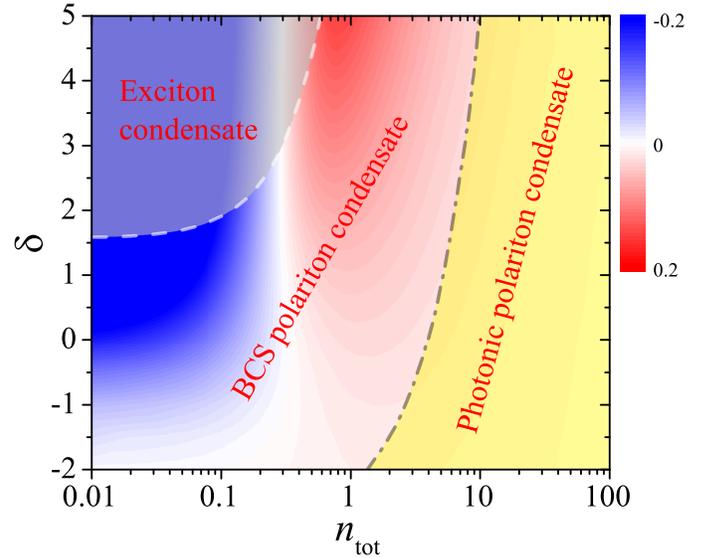}\caption{\label{fig1_PhaseDiagram} Zero-temperature phase diagram as functions
of the total carrier density $n_{\textrm{tot}}$ (in units of $n_{\textrm{mott}}=a_{B}^{-2}$)
and the cavity detuning $\delta$ (in units of $E_{B}$), at the light-matter
coupling $\Omega=0.8E_{B}$. The color shows the difference in density
fractions of fermionic quasi-particles and excitons. The phase diagram
is divided into three parts by the dashed and dot-dashed lines, which
indicate $10\%$ and $90\%$ photonic fractions, respectively. A BCS
polariton condensate forms in between, where Bogoliubov quasi-particles
and excitons co-exist. The boundaries between different phases should
be understood as crossover, rather than phase transition.}
\end{figure}

Microscopic theoretical description of the evolution from an exciton-polariton
BEC to a BCS polariton condensate, the so-called BEC-BCS crossover,
is highly non-trivial due to several reasons: (i) The inter-particle
Coulomb interactions are strong and long-range; (ii) The coupling
between light and matter can also be non-perturbative in the \emph{ultra-strong}
coupling regime achieved so far in experiment \cite{Kockum2019},
where the characteristic Rabi coupling $\Omega$ can be comparable
to the binding energy $E_{B}$ of excitons; (iii) Electron-hole-photon
systems are often confined in single-layer quantum well and therefore
are two-dimensional in character, where both quantum and thermal fluctuations
are significant \cite{Loktev2001}; and (iv) The systems are inherently
non-equilibrium. Continuous pumping is necessary to compensate the
loss in carriers, in order to keep a steady state \cite{Deng2010,Carusotto2013}.
The first theoretical study of the challenging problem of BEC-BCS
crossover with exciton-polaritons was provided by Littlewood and his
co-workers \cite{Eastham2001,Keeling2004,Keeling2005}, by treating
localized excitons as a collection of two-level systems. A more formal
treatment was later developed by Kamide \textit{et al.} \cite{Kamide2010}
and Byrnes \textit{et al.} \cite{Byrnes2010}, by using the celebrated
BCS variational wave-functions. While the mean-field BCS wave-function
permits a description of fermionic quasi-particles and of condensed
photons and gives a useful equilibrium phase diagram \cite{Kamide2010,Xue2016},
strictly speaking, it does not provide information on the key ingredient
of exciton-polariton quasi-particles, which should be obtained by
taking into account quantum fluctuations beyond mean-field. More seriously,
in the limit of strong coupling mean-field framework may simply break
down in two dimensions, even at the qualitative level, as shown by
the recent investigation of BEC-BCS crossover in a two-dimensional
strongly interacting Fermi gas \cite{Makhalov2014,He2015}.

The purpose of this Rapid Communication is to deepen our understanding
of polariton quasi-particles in strongly interacting BCS polariton
condensates, by adopting a zero-temperature Gaussian pair fluctuation
(GPF) theory \cite{He2015,Ohashi2003,Hu2006,Hu2007,Diener2008}. A
more reliable phase diagram at thermal equilibrium is then determined
(see Fig. 1). The inclusion of the crucial quantum fluctuations comes
at a price. We need to replace the long-range Coulomb force between
electrons and holes by a \emph{short-range} contact interaction. This
seems to be a reasonable approximation in the dilute exciton limit,
where the strength of the contact interaction is tuned to reproduce
the exact binding energy of excitons. Towards the high-density regime,
the approximation may get improved, considering the gradual screen
of Coulomb interactions and the increasingly dominant role played
by the light-matter coupling. We note that, a contact electron-hole
interaction was previously used in the mean-field treatment, which
leads to more or less similar results as the long-range Coulomb interaction
\cite{Yamaguchi2012}. In this work we also consider the equilibrium
situation only, as inspired by the recent experimental demonstration
of a \emph{thermal }equilibrium exciton-polariton BEC in a high-quality
microcavity \cite{Sun2017}. The extension of our work to include
dissipations is possible, following the standard Keldysh technique
\cite{Yamaguchi2013,Yamaguchi2015,Hanai2018}.

Our key results can be briefly summarized as follows. First, as shown
in Fig. 1, in the ultra-strong coupling regime we observe an unexpectedly
wide phase space for BCS polariton condensate with strong-correlated
$e$-$h$ pairs coupled to photons. The window for conventional exciton-polariton
BEC shrinks due to ultra-strong light-matter coupling. A BCS superfluid
of $e$-$h$ pairs only, anticipated in the previous studies \cite{Kamide2010},
is also not favorable. Secondly, at high carrier density where photons
become dominant, the system remains strongly coupled or correlated,
as indicated by a large fraction of non-condensed photons. It is then
better characterized as a photonic polariton condensate. This picture
is consistent with earlier theoretical \cite{Ishida2014} and experimental
investigations \cite{Horikiri2016,Horikiri2017}. Finally, with increasing
carrier density, we find that both lower and upper polariton energies
of the BCS polariton condensate get strongly renormalized. In particular,
the upper polariton branch shifts up significantly. This finding may
provide a qualitative explanation for the puzzling observation of
a high-energy side-peak in emission spectra at high densities \cite{Horikiri2016}.

\textit{Microcavity electron-hole-photon systems. }We start by considering
the following approximate model Hamiltonian in two dimensions with
area $\mathcal{S}=1$ \cite{Ohashi2003,Yamaguchi2012,Hanai2018}:
\begin{eqnarray}
\mathcal{H} & = & \sum_{\mathbf{k}\sigma}\xi_{\mathbf{k}}c_{\mathbf{k}\sigma}^{\dagger}c_{\mathbf{k}\sigma}+u_{0}\sum_{\mathbf{k}\mathbf{k}'\mathbf{q}}c_{\frac{\mathbf{q}}{2}+\mathbf{k}e}^{\dagger}c_{\frac{\mathbf{q}}{2}-\mathbf{k}h}^{\dagger}c_{\frac{\mathbf{q}}{2}-\mathbf{k}'h}c_{\frac{\mathbf{q}}{2}+\mathbf{k}'e}\nonumber \\
 &  & +g_{0}\sum_{\mathbf{kq}}\left(\phi_{\mathbf{q}}^{\dagger}c_{\frac{\mathbf{q}}{2}-\mathbf{k}h}c_{\frac{\mathbf{q}}{2}+\mathbf{k}e}+\textrm{h.c.}\right)+\sum_{\mathbf{q}}\omega_{\mathbf{q}}\phi_{\mathbf{q}}^{\dagger}\phi_{\mathbf{q}},\label{eq:hami}
\end{eqnarray}
where $c_{\mathbf{k}\sigma}$ are the annihilation operators of electrons
($\sigma=e$) and holes ($\sigma=h$) with dispersion $\xi_{\mathbf{k}}\equiv\hbar^{2}\mathbf{k}^{2}/(2m_{\textrm{eh}})-\mu_{\textrm{eh}}/2$,
and $\phi_{\mathbf{q}}$ are the annihilation operators of photons
with dispersion $\omega_{\mathbf{q}}\equiv\hbar^{2}\mathbf{q}^{2}/(2m_{\textrm{ph}})+\delta_{0}-\mu_{\textrm{ph}}$.
In both dispersion relations, the chemical potentials ($\mu_{\textrm{eh}}$
for electrons/holes and $\mu_{\textrm{ph}}$ for photons) are measured
from the band gap $E_{\textrm{g}}$ of the semiconductor quantum well
and are the same in equilibrium. For convenience, we have assumed
the same mass $m_{\textrm{eh}}\simeq0.068m_{0}$ for electrons and
holes (where $m_{0}$ is the mass of free electrons) \cite{Deng2010,Hanai2018}.
The mass of photons due to quantum well confinement is instead much
smaller and we set $m_{\textrm{ph}}\simeq3\times10^{-5}m_{0}$ \cite{Deng2010,Hanai2018}.
The detuning of the microcavity with respect to the band edge is denoted
by $\delta_{0}$ and is to be replaced with $\delta=E_{\textrm{cav}}-E_{\textrm{g}}$
upon renormalization \cite{SM}. The second and third terms in the
Hamiltonian describe the electron-hole attraction and light-matter
coupling, respectively, with \emph{bare} interaction strengths ($u_{0}$
and $g_{0}$), which are to be renormalized to reproduce the observable
exciton binding energy $E_{B}$ and Rabi coupling $\Omega$. We refer
to Supplemental Material \cite{SM} for two-body physics and the detailed
procedure of renormalization. From now on, we always use the renormalized
parameters $u$, $g$ and $\delta$ in equations.

The Hamiltonian Eq. (\ref{eq:hami}) provides a reasonable \emph{minimal}
model of electron-hole-photon systems in microcavity \cite{Yamaguchi2012,Hanai2018}.
As the light-matter coupling generates an effective attraction between
electrons and holes \cite{Ohashi2002,Citrin2003,Liu2015}, which eventually
dominates over the direct Coulomb interaction at high carrier density,
we anticipate that the replacement of Coulomb interaction in terms
of a short-range contact interaction only brings corrections at the
\emph{quantitative} level. We note that the model Hamiltonian also
describes a strongly interacting Fermi gas of ultracold atoms near
a narrow Feshbach resonance \cite{Ohashi2003}, which unfortunately
suffers from severe atom loss \cite{Hazlett2012}. A near-equilibrium
high-density exciton-polariton therefore realizes an interesting example
to explore the many-body physics with large effective range of interactions
\cite{Hazlett2012,Hu2019,Wu2019}.

\begin{figure}[t]
\centering{}\includegraphics[width=0.5\textwidth]{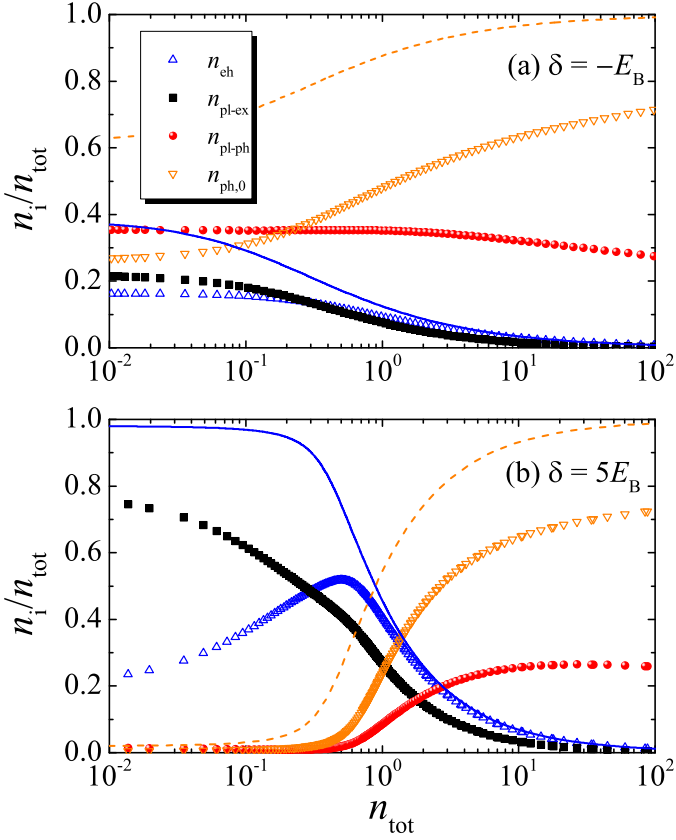}\caption{\label{fig2_DensityCurve} Density fraction of different carriers
as a function of the total carrier density (in units of $a_{B}^{-2}$),
at the Rabi coupling $\Omega=0.8E_{B}$ and the detuning $\delta=-E_{B}$
(a) and $\delta=5E_{B}$ (b). For comparison, we also show the mean-field
results for fermionic quasi-particles $n_{\textrm{eh}}$ and condensed
photons $n_{\textrm{ph,0}}$, using blue solid lines and orange dashed
lines, respectively. }
\end{figure}

\textit{GPF framework}. The use of a short-range electron-hole attraction
allows us to understand the crucial role of quantum fluctuations in
exciton-polariton condensates within the reliable GPF theory \cite{He2015,Ohashi2003,Hu2006,Hu2007,Diener2008,footnoteGPF}.
It can be mostly easily derived with the help of the functional path-integral
approach, as given in Supplemental Material \cite{SM}. Here, we briefly
review the key steps. By using the Hubbard\textendash Stratonovich
transformation, we first introduce a pairing field $\varphi_{\mathbf{q}}$
to decouple the electron-hole interaction term and integrate out the
fermionic fields $c_{\mathbf{k}\sigma}$. This leads to an effective
action for the pairing field $\varphi_{\mathbf{q}}$ and photon field
$\phi_{\mathbf{q}}$. The saddle-point solution of both fields (more
precisely, their superposition) then gives rise to the mean-field
thermodynamic potential at $T=0$,
\begin{equation}
\Omega_{\textrm{MF}}=-\frac{\Delta^{2}}{u_{\textrm{eff}}}+\sum_{\mathbf{k}}\left[\xi_{\mathbf{k}}-E_{\mathbf{k}}+\frac{\Delta^{2}}{\hbar^{2}\mathbf{k}^{2}/m_{\textrm{eh}}+\varepsilon_{0}}\right],\label{eq:OmegaMF}
\end{equation}
where $\Delta$ is an order parameter to be determined by the gap
equation $\partial\Omega_{\textrm{MF}}/\partial\Delta=0$, $u_{\textrm{eff}}\equiv u+g^{2}/(\mu_{\textrm{ph}}-\delta)$
is an effective interaction incorporating the photon-mediated attraction
\cite{Ohashi2002,Liu2015}, $E_{\mathbf{k}}\equiv\sqrt{\xi_{\mathbf{k}}^{2}+\Delta^{2}}$
is the dispersion relation for fermionic quasi-particles, and $\varepsilon_{0}\ll E_{B}$
is an unimportant energy scale to regularize the logarithmic infrared
divergence in two dimensions \cite{SM}. According to the standard
BCS theory, the mean-field thermodynamic potential in Eq. (\ref{eq:OmegaMF})
describes the\emph{ fermionic} Bogoliubov quasi-particles, as well
as the \emph{condensed} photons. To go beyond mean-field, we consider
the bilinear Gaussian fluctuations around the saddle point, which
may physically be interpreted as \emph{polaritons}. Integrating out
these fluctuations, we obtain the GPF thermodynamic potential from
quantum fluctuations,
\begin{equation}
\Omega_{\textrm{GPF}}=\frac{1}{2}k_{B}T\sum_{\mathcal{Q}}\ln\left[M_{11}M_{22}-M_{12}M_{21}\right]e^{i\nu_{n}0^{+}},\label{eq:OmegaGPF}
\end{equation}
where $\mathcal{Q\equiv}(\mathbf{q},i\nu_{n})$ with bosonic Matsubara
frequencies $\nu_{n}=2\pi nk_{B}T$ ($n\in\mathbb{Z}$) and $M_{ij}(\mathcal{Q})$
($i,j=1,2$) are the matrix elements of the 2 by 2 polariton Green
function,\begin{widetext}

\begin{eqnarray}
M_{11}\left(\mathcal{Q}\right)=M_{22}\left(-\mathcal{Q}\right) & = & \left[-\frac{1}{u_{\textrm{eff}}\left(\mathcal{Q}\right)}+\frac{1}{u_{\textrm{eff}}}\right]+\sum_{\mathbf{k}}\left[\frac{u_{+}^{2}u_{-}^{2}}{i\nu_{n}-E_{+}-E_{-}}-\frac{v_{+}^{2}v_{-}^{2}}{i\nu_{n}+E_{+}+E_{-}}+\frac{1}{2E_{\mathbf{k}}}\right],\\
M_{12}\left(\mathcal{Q}\right)=M_{21}\left(+\mathcal{Q}\right) & = & \sum_{\mathbf{k}}u_{+}v_{+}u_{-}v_{-}\left[\frac{1}{i\nu_{n}-E_{+}-E_{-}}-\frac{1}{i\nu_{n}+E_{+}+E_{-}}\right].
\end{eqnarray}
\end{widetext}Here, we define the short-hand notations, $E_{\pm}\equiv E_{\mathbf{k}\pm\mathbf{q}/2}$,
$u_{\pm}^{2}=(1+\xi_{\mathbf{k}\pm\mathbf{q}/2}/E_{\mathbf{k}\pm\mathbf{q}/2})/2$,
$v_{\pm}^{2}=1-u_{\pm}^{2}$, and a momentum- and frequency-dependent
effective interaction strength $u_{\textrm{eff}}(\mathcal{Q})\equiv u+g^{2}/[i\nu_{n}-\hbar^{2}\mathbf{q}^{2}/(2m_{\textrm{ph}})+\mu_{\textrm{ph}}-\delta]$.
For given $\mu_{\textrm{eh}}=\mu_{\textrm{ph}}=\mu$, we calculate
the total thermodynamic potential $\Omega_{\textrm{tot}}=\Omega_{\textrm{MF}}+\Omega_{\textrm{GPF}}$
and then determine the total density of carriers, $n_{\textrm{tot}}=-\partial\Omega_{\textrm{tot}}/\partial\mu$.

\textit{Nature of carriers and phase diagrams}. According to the structure
of mean-field and GPF thermodynamic potentials, $\Omega_{\textrm{MF}}(\mu_{\textrm{eh}},\mu_{\textrm{ph}})$
and $\Omega_{\textrm{GPF}}(\mu_{\textrm{eh}},\mu_{\textrm{ph}})$,
we may distinguish \emph{four} types of carriers:
\begin{eqnarray}
n_{\textrm{eh}} & = & -\frac{\partial\Omega_{\textrm{MF}}}{\partial\mu_{\textrm{eh}}}=\frac{1}{2}\sum_{\mathbf{k}}\left(1-\frac{\xi_{\mathbf{k}}}{E_{\mathbf{k}}}\right),\\
n_{\textrm{ph,0}} & = & -\frac{\partial\Omega_{\textrm{MF}}}{\partial\mu_{\textrm{ph}}}=\left(\frac{g}{\delta-\mu_{\textrm{ph}}}\right)^{2}\frac{\Delta^{2}}{u_{\textrm{eff}}^{2}},
\end{eqnarray}
$n_{\textrm{pl-ex}}=-\partial\Omega_{\textrm{GPF}}/\partial\mu_{\textrm{eh}}$
and $n_{\textrm{pl-ph}}=-\partial\Omega_{\textrm{GPF}}/\partial\mu_{\textrm{ph}}$.
The former two are the fermionic quasi-particles \cite{footnoteFermiQP}
and condensed photons, while the latter two could be understood as
the excitonic and photonic parts of the exciton-polariton quasi-particles.
In the previous studies \cite{Kamide2010,Byrnes2010,Yamaguchi2012},
only the numbers of fermionic quasi-particles and condensed photons
are accessible using a BCS variational wave-function.

In Fig. \ref{fig2_DensityCurve}, we show the density fraction of
different carriers at zero cavity detuning (a) and at a large detuning
(b) in different symbols, as a function of the total carrier density
at an ultra-strong light-matter coupling $\Omega=0.8E_{B}$. To compare
with earlier results \cite{Yamaguchi2012}, we plot also the mean-field
prediction for $n_{\textrm{eh}}$ and $n_{\textrm{ph},0}$ with the
solid and dashed lines, respectively. At zero detuning in Fig. \ref{fig2_DensityCurve}(a),
the photon energy $E_{\textrm{cav}}$ is equal to the exciton energy
$E_{\textrm{X}}=-E_{B}$; both of them are measured from the band
gap energy $E_{\textrm{g}}$. In the limit of low carrier density,
in sharp contrast to the conventional exciton-polariton picture (i.e.,
a polariton consists of half-exciton and half-photon), we find that
all the four carriers are notably populated, suggesting the appearance
of a BCS-like polariton condensate, where the fermionic degree of
freedom is important and $n_{\textrm{eh}}\simeq n_{\textrm{pl-ex}}$.
The significant population of fermionic quasi-particles $n_{\textrm{eh}}$
can be understood from the ultra-strong light-matter coupling: as
a result of $\Omega\sim E_{B}$, the internal fermionic degree of
freedom of excitons is no longer frozen and can directly manifest
itself in the few-body wave-function, as discussed in a recent microscopic
calculation \cite{Levinsen2019}. The low-density situation at a large
cavity detuning in Fig. 2(b) seems to be different. Although $n_{\textrm{eh}}$
is still non-negligible, photons are now depleted due to their large
energy up-shift. The density of excitons $n_{\textrm{pl-ex}}$ is
dominant, much larger than $n_{\textrm{eh}}$. The system is then
better viewed as an exciton condensate without photons. This picture
becomes less accurate when the total carrier density increases, since
the photon-mediated attraction plays an increasingly important role
and drives the system into the BEC-BCS crossover regime with a notable
photon fraction \cite{Ohashi2003}. As a consequence, close to the
Mott transition density $n_{\textrm{tot}}\sim a_{B}^{-2}$, we find
again $n_{\textrm{eh}}\sim n_{\textrm{pl-ex}}$ and hence a strongly
interacting BCS polariton condensate. Finally, at sufficiently large
carrier density $n_{\textrm{tot}}\gg a_{B}^{-2}$, photons with density
$n_{\textrm{ph}}=n_{\textrm{ph,0}}+n_{\textrm{pl-ph}}$ becomes dominant
at both zero and large detunings due to the Bose enhancement \cite{Yamaguchi2012},
and forms a photonic condensate. This observation is consistent with
earlier mean-field predictions \cite{Kamide2010,Byrnes2010,Yamaguchi2012}.
However, our beyond-mean-field theory is able to reveal an important
new feature: not all the photons are in the condensate and the density
of \emph{non-condensed} photons (i.e., $n_{\textrm{pl-ph}}$) is always
significant. To characterize this strong-coupling state, the system
could be better termed as a photonic polariton condensate.

\begin{figure}[t]
\begin{centering}
\includegraphics[width=0.5\textwidth]{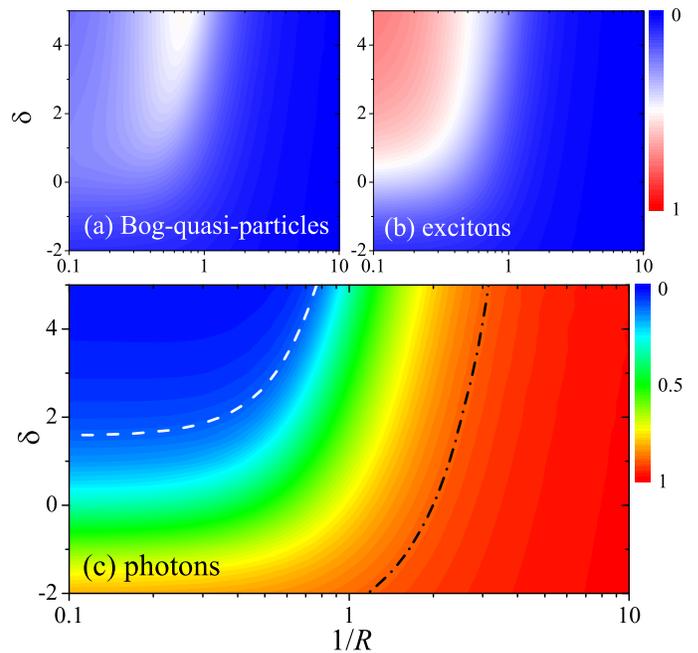} 
\par\end{centering}
\centering{}\caption{\label{fig3_DensityContour} Contour plot of the density fractions
of fermionic Bogoliubov quasi-particles (a), excitons (b), and photons
(c), as functions of the inverse mean particle distance $1/R=n_{\textrm{tot}}^{1/2}$
(in units of $a_{B}^{-1}$) and the detuning $\delta$ (in units of
$E_{B}$). The dashed and dot-dashed lines in (c) show $10\%$ and
$90\%$ photonic fractions, respectively. }
\end{figure}

We now examine the detuning dependence of the different carrier fractions,
as presented in the contour plot Fig. \ref{fig3_DensityContour} for
fermionic quasiparticles $n_{\textrm{eh}}$ (a), excitonic part of
polaritons $n_{\textrm{pl-ex}}$ (b), and total photons $n_{\textrm{ph}}$(c).
Here, the dimensionless ratio $R=n_{\textrm{tot}}^{-1/2}/a_{B}$ measures
the average distance between carriers relative to the Bohr radius
of excitons. From Fig. \ref{fig3_DensityContour}(c), we can see clearly
that the total photon density increases monotonically with decreasing
cavity detuning $\delta$ or mean carrier separation $R$. The photon-rare
and photon-rich areas may be qualitatively characterized by the dashed
and dot-dashed lines, which denote $10\%$ and $90\%$ photonic fractions,
respectively. As the BCS polariton condensate, which is of major interest
here, has approximately equal densities of fermionic quasi-particles
($n_{\textrm{eh}}$) and excitons ($n_{\textrm{pl-ex}}$), we calculate
the difference of these two density fractions using Fig. \ref{fig3_DensityContour}(a)
and \ref{fig3_DensityContour}(b), and show the result in Fig. \ref{fig1_PhaseDiagram}.
Masking further the photon-rare and photon-rich areas on it, we then
obtain a phase diagram.

We may identify three phases. In the top left corner of the phase
diagram (Fig. \ref{fig1_PhaseDiagram}), photons are basically absent
and carriers are dominated by excitons, forming an exciton condensate.
On the contrary, in the bottom right corner of the figure, carriers
are mostly photons, giving rise to a photonic polariton condensate,
as discussed earlier. In between, it is interesting to see a large
phase space for the BCS polariton condensate, with nearly equal numbers
of fermionic quasi-particles and excitons, both of which couple to
photons. This is the main result of our work. We note that, compared
with earlier mean-field predictions \cite{Kamide2010}, we do not
find a BCS superfluid with $e$-$h$ pairs only. There, the $e$-$h$
BCS state is indirectly determined from the mean-field pair wave-function,
which shows a peak in momentum space \cite{Kamide2010,Byrnes2010}.
Our way of comparing the numbers of fermionic quasi-particles and
excitons seems to be more direct and reasonable. We also note that,
the large phase space of BCS polariton condensate is a consequence
of the ultra-strong light-matter coupling. If we decrease the coupling
by a factor of 10 (i.e., \emph{strong} light-matter coupling regime),
in the low carrier density limit we instead observe a conventional
exciton-polariton condensate and the window for a BCS polariton condensate
becomes considerably narrower \cite{SM}.

\begin{figure}[t]
\begin{centering}
\includegraphics[width=0.5\textwidth]{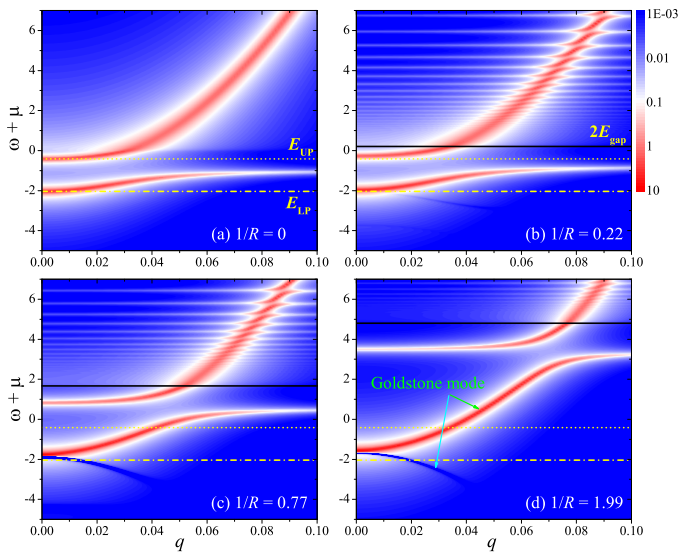} 
\par\end{centering}
\centering{}\caption{\label{fig4_SpectralFunction} Absorption coefficient $\mathcal{A}(q,\omega)$
(in arbitrary units and in logarithmic scale) at different inverse
mean carrier separations $1/R=0$ (a), $0.22a_{B}^{-1}$ (b), $0.77a_{B}^{-1}$
(c) and $1.99a_{B}^{-1}$ (d), and at zero detuning $\delta=-E_{B}$
and the Rabi coupling $\Omega=0.8E_{B}$. The wave-vector $q$ is
in units of $a_{B}^{-1}$ and the frequency $\omega$ is in units
of $E_{B}$. The yellow dot-dashed lines and dotted lines show the
lower- and upper-polariton energies in the zero-density limit, respectively.
The solid black lines is the threshold of two-particle continuum,
$E_{\textrm{th}}=2E_{\textrm{gap}}$, where $E_{\textrm{gap}}$ is
the energy gap for single-particle excitations. For better illustration,
a spectral broadening of $0.05E_{B}$ is used.}
\end{figure}

\textit{Photonic spectra}. How to experimentally probe the intriguing
strongly-correlated BCS polariton condensate? A unique advantage of
an electron-hole-photon system is that its photoluminescence provides
direct information of underlying many-body physics. Therefore, we
consider the photonic spectral function $\mathcal{A}(\mathbf{q},\omega)\equiv-(1/\pi)\textrm{Im}\mathcal{D}(\mathbf{q},i\nu_{n}\rightarrow\omega+i0^{+})$,
which experimentally corresponds to the absorption coefficient. Here,
the Green function of photons $\mathcal{D}(\mathcal{Q})=[\textrm{diag}\{D_{0}(\mathcal{Q}),D_{0}(-\mathcal{Q})\}-\Sigma{}_{\textrm{ph}}(\mathcal{Q})]^{-1}$
can be calculated using the free Green function $D_{0}(Q)=[i\nu_{n}-\hbar^{2}\mathbf{q}^{2}/(2m_{\textrm{ph}})+\mu_{\textrm{ph}}-\delta+g^{2}/u]^{-1}$
and the photon self-energy
\begin{equation}
\Sigma_{\textrm{ph}}\left(\mathcal{Q}\right)=\frac{g^{2}}{u^{2}}\left[\begin{array}{cc}
B\left(\mathcal{Q}\right)-M_{11} & M_{12}\\
M_{21} & B\left(-\mathcal{Q}\right)-M_{22}
\end{array}\right]^{-1}
\end{equation}
with $B(\mathcal{Q})\equiv u^{-1}-u_{\textrm{eff}}^{-1}(\mathcal{Q})$
\cite{SM}.

In Fig. \ref{fig4_SpectralFunction}, we report the absorption coefficient
vs momentum and energy at zero detuning $\delta=-E_{B}$ and at $\Omega=0.8E_{B}$,
with increasing carrier density from the dilute limit (a) to the Mott
transition (d). There are always two bright branches, exhibiting an
avoided-crossing at certain momentum. The lower branch can be identified
as the \emph{gapless} Goldstone mode, since the excitation energy
$\omega$ goes to \emph{zero} as momentum $q$ vanishes \cite{SM,footnotePL}.
It stays within the energy gap for single-particle excitations (i.e.,
$\left|\omega\right|<2E_{\textrm{gap}}$, the threshold for two-particle
continuum) and therefore should behaves like a delta-peak if we do
not consider any spectral broadening. On the other hand, the upper
branch can enter the two-particle continuum. As a result, this branch
has an intrinsic spectral width, as indicated by a lot of additional
irregular wave-like structures at $\omega>2E_{\textrm{gap}}$. In
the low-density limit, we may also recognize the two branches as the
lower- and upper-polariton branches, respectively. In such a understanding,
as the carrier density increases, the upper-polariton energy gets
strongly renormalized. 

\begin{figure}[t]
\begin{centering}
\includegraphics[width=0.5\textwidth]{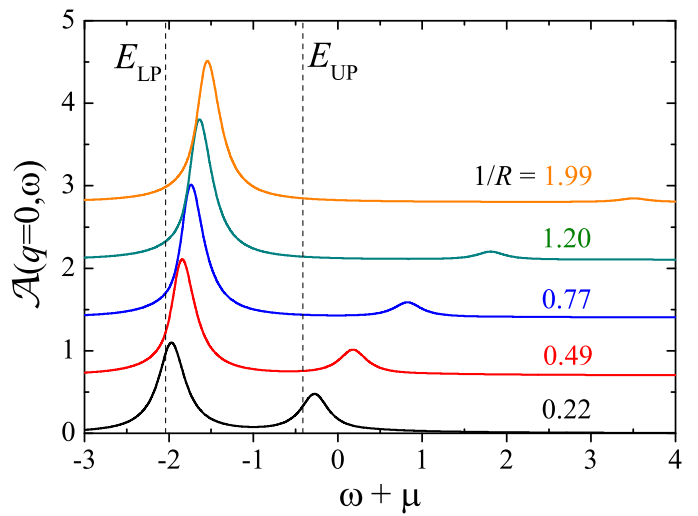} 
\par\end{centering}
\centering{}\caption{\label{fig5_SidePeak} $\mathcal{A}(\mathbf{q}=0,\omega)$ (in arbitrary
units) at different inverse mean particle distances as indicated.
Here, the cavity detuning $\delta=-E_{B}$ and the Rabi coupling $\Omega=0.8E_{B}$.
The two dashed lines show the lower-polariton and upper-polariton
energies in the zero-density limit. The frequency $\omega$ is in
units of $E_{B}$. We have used a spectral broadening of $0.2E_{B}$.}
\end{figure}

This can be seen more clearly from the carrier density dependence
of the spectral function at zero momentum $\mathcal{A}(\mathbf{q}=0,\omega)$,
as shown in Fig. \ref{fig5_SidePeak}. Here, the high-energy peak
show a rapid blue-shift with increasing carrier density or with decreasing
mean separation $R$. It is interesting that such a shift is in a
qualitative agreement with the recent observation of a high-energy
side-peak in photoluminescence in high carrier density regime, which
turns out to be difficult to understand theoretically in terms of
Mollow's triplet \cite{Horikiri2016}. We also find that the energy
of the lower polaritons shifts up gradually towards the cavity photon
energy $E_{\textrm{cav}}=-E_{B}$. This blue-shift is well-known and
has been phenomenologically attributed to the polariton-polariton
interaction \cite{Deng2010}. Our result provides a useful explanation
based on the microscopic model calculation.

\textit{Conclusions}. We have presented a microscopic, beyond-mean-field
study of a novel strongly interacting exciton-polariton condensate
at high density, the so-called BCS polariton condensate. In our theoretical
modeling, polariton quasi-particles emerge as quantum fluctuations
and their direct characterization allows us to determine the phase
diagram with an improved accuracy. We have found that the BCS polariton
phase occupies a significant phase space at ultra-strong light-matter
coupling. We have predicted the photoluminescence spectra, which might
be useful for its eventual experimental confirmation.
\begin{acknowledgments}
This research was supported by the Australian Research Council's (ARC)
Discovery Program, Grant No. DP170104008 (H.H.), Grant No. FT140100003
(X.-J.L), and Grant No. DP180102018 (X.-J.L).
\end{acknowledgments}


\begin{thebibliography}{10}
\bibitem{footnoteFermiQP}In the normal state ($\Delta=0$), the expression
for fermionic quasi-particles reduces to $n_{\textrm{eh}}=\sum_{\mathbf{k}}1/[1+e^{\xi_{\mathbf{k}}/(k_{B}T)}]$,
which is the half of the density of a non-interacting Fermi gas of
electrons and holes. In the superfluid phase, we may naively identify
$n_{\textrm{eh}}$ as the number of weakly-bound, strongly-overlapping
$e$-$h$ pairs. This interpretation is not accurate in the BCS limit,
where the number of $e$-$h$ pairs goes to zero.
\bibitem{footnotePL}This may not be so obvious in Fig. \ref{fig4_SpectralFunction}.
We note that, in our calculations the excitation energy $\omega$
in the photonic spectral function $\mathcal{A}(\mathbf{q},\omega)$
is measured from the chemical potential $\mu$. To show the excitation
energy as observed in the experiments, we need to plot $\mathcal{A}(\mathbf{q},\omega)$
as a function of $\omega+\mu$.
\end{thebibliography}
\end{document}